# Modeling of eddy current distribution in the SST-1 tokamak

Amit K. Singh, Santanu Banerjee*, I. Bandyopadhyay, Deepti Sharma, S. K. Jha,
R. Srinivasan, D. Raju, M.V. Gopalakrishna, and the SST-1 team

*Institute for Plasma Research, HBNI, Bhat, Gandhinagar 382428, Gujarat, India*

*Email of corresponding author: sbanerje@ipr.res.in

## ABSTRACT

The time varying currents in the Ohmic transformer in the SST-1 tokamak induce voltages that drive large eddy currents in the passive structures like the vacuum vessel and cryostat. Since the vacuum vessel and the cryostat are toroidally continuous without electrical breaks in SST-1, this leads to a shielding effect on the flux penetrating the vacuum vessel. This reduces the magnitude of the loop voltage seen by the plasma as also delays its buildup. Also the induced currents alter the null location of magnetic field. This will have serious implications on the plasma breakdown and startup and corrective measures may be required in case of an insufficient loop voltage or an improper null. Further, the eddy currents distribution will be vital for the plasma equilibrium and need to be considered while reconstructing the equilibrium. Evolution of the toroidal eddy currents in SST-1 passive structures has been studied using a toroidal-filament model. The model calculations are compared with the measured signals in the magnetic diagnostics like the toroidal flux loops and magnetic pick-up coils installed on the SST-1.

Keywords: filament model, magnetics, flux loops

## I. Introduction

Modeling of the induced eddy current distribution in the passive conducting structures, like the vacuum vessel and the cryostat in a superconducting tokamak is a complex task. This is indispensable for a number of reasons like, the plasma current start-up, position control, disruptions and their aftermath, equilibrium reconstruction etc. In present day superconducting tokamaks, there are severe design constraints and those make the eddy current contributions rather significant. It is mechanically impractical to construct a tokamak vacuum vessel to make the toroidal eddy current flow negligible [1]. This problem has worsened in the superconducting tokamaks with the advent of the cryostat assembly. The eddy current flow in the passive elements inhibits the penetration of Ohmic flux and vertical field making the plasma start-up and burn-through phases quite complicated. It also plays an important role in stabilizing the positional instability of the plasma [2-3] and thereby renders the position control extremely difficult. The magnetic fields produced by the eddy currents could generate error fields that may give rise to islands at the rational surfaces or introduce chaos to the magnetic field lines, enhancing anomalous cross-field transport [4]. Further, there may be additional poloidal field (PF) coils on the tokamak which are sitting idle (typically shortened with a resistor ~ 0.3 Ω) for a particular discharge scenario. The induced voltages can pose serious threat to the insulation of these coils to the extent of damaging the insulation altogether. Hence, the tokamak assembly and operation personnel need to know the magnitude and spatial distribution of these currents. Given the importance and complexity of the problem, the available literature on the eddy current modeling in tokamaks is still far from adequate. Our effort in this paper is to develop a near comprehensive understanding of the eddy current distribution in a tokamak and its effect on the magnetic measurements.

The eddy current distribution is generally modelled in two types of approaches. In the first method, the passive conducting structures in the tokamak are replaced by toroidally symmetric passive filaments [2]. This approach will be referred as the filament model henceforth. The other approach is to use the finite element method (FEM) [3,5] to model the structures. The filament model is generally used in plasma simulations and is relatively easier to handle. The caveat of this approach is that, it is virtually impossible to analyze the effects of eddy currents in the complicated mechanical structures in a tokamak with this model. Hence, the filament model can be used efficiently unless the 3D effects of eddy currents are significant for a given application. On the other hand, precise modeling of the structures and their material properties are possible using FEM. FEM is computationally expensive since the modeling is in three dimensions (3D). Hence, it is essential to study the reliability and robustness of the filament model for a given tokamak and for the concerned operation scenarios and model application.

In this paper we describe the modelling of the eddy current distribution in the cryostat, vacuum vessel and other passive structures on the Steady-state Superconducting Tokamak, SST-1 [6-8] using a filament model. View ports and other complex structures on the tokamak are either neglected or simplified as toroidally symmetric filaments to maintain the brevity of the analysis. This approximation makes it readily usable for a wide variety of purposes. The estimated eddy current distribution is validated against the experimentally measured signals in various magnetic diagnostics like the toroidal flux loops and magnetic pick-up (position probe) coils.

The paper is organized as follows: the SST-1 machine is described briefly in section II. The filament model is stated in section III. Results of the experimental validation are shown in section IV. Finally the results and the outstanding issues are discussed in section V.

## II. The SST-1 machine

The SST-1 tokamak is a large aspect ratio tokamak, configured to run double null diverted plasmas with significant elongation $\kappa$ and triangularity $\delta$. The machine has a major radius of 1.1 m, a minor radius of 0.20 m, maximum toroidal magnetic field ($B_t$) of 3.0 T at the plasma center [6]. Elongated plasma with elongation in the range 1.7-1.9 and triangularity in the range 0.4-0.7 can be produced. The magnet system comprises the TF coil system, the PF coil system, the Ohmic transformer, the vertical field coils and the vertical position control coils. 6 PF coil pairs (except PF1) are installed for realizing various operational scenarios. PF1 to PF5 are superconducting coils while PF6 is a normal conductor coil. An Ohmic transformer, comprised of a central solenoid (TR1) and two pairs of compensation coils (TR2 and TR3) will, therefore, be used for plasma startup and initial current ramp-up. These coils are made from hollow copper conductors. The TR2 and TR3 coils minimize the magnetic field produced by TR1 in the plasma to values of less than 10 G at full flux storage. The transformer has a storage flux of 1.4 V s and can be used for producing circular plasma with currents up to 100 kA for almost 1 s. A pair of vertical field (BV) coils keeps this circular plasma in equilibrium during the initial phase. This pair of normal conductor coils is required to establish an equilibrium field on a fast timescale during plasma startup and current ramp-up in the initial stage. It can be noted that, limiter bound plasma configuration of circular cross-section, taken at $B_t$ = 1.5 T are considered for this work. Physical parameters, like radial (*R*) and vertical (*Z*) position from the machine center, radial and vertical spans (*dR* and *dZ*), number of turns ($N_{turn}$), inductance (*L*) and resistance (*Res*), of all the coils in SST-1 are given in table 1. Entries numbered as 1-7 in table 1 are for the components of the central solenoid (TR# coils), entries 8-9 are for the BV coil pair, entries 10-18 are for the PF coil system and entry number 19 is for a representative toroidal passive filament of the vacuum vessel near the mid-plane. It can be noted that the passive components are divided into several filaments as stated in the following section. Only one of those filaments, representing a constituent part of the vacuum vessel, is given as entry number 19 in table 1.

## III. The filament model

A code, called INDUCT (abbreviation from INDUCTance calculation) is developed on the basis of the filamentary model and is used for the eddy current calculations. In SST-1 the vacuum vessel and cryostat are conductors with large poloidal cross-section. While modeling, these passive structures are broken up into a large number (~1000) of co-axial toroidal current carrying filaments of circular cross-section. The radial and vertical spans of each toroidal passive conductor segment (*dR* and *dZ*) is broken up into a set of toridal filaments as shown in Fig.1. The inductance matrix for this large set of toroidal current carrying conductors is calculated using the standard Green functions. The total number of filaments required to represent the passive structures are optimized by slowly increasing the number until the elements of the inductance matrix are saturated ($\sim 2\times10^{-7}$ H). The induced currents are evaluated by solving a set first order ordinary differential equations (ODEs) for the circuit equations. The ODEs are solved using the Runge-Kutta method (RK4). The step size in RK4 is optimized at 100 $\mu$s to achieve a tolerance level of $\sim 10^{-8}$ A for the evaluated induced currents in the range of ~100 A – 1 kA. It can be noted that, tolerance level achieved depends on the optimized step size and does not depend on the number of equations involved in RK4. The calculated inductance matrix is validated for test cases with the code for calculating the electromagnetic field, force and inductance (EFFI) in coil systems of arbitrary geometry [9]. The contribution of TF coils (if any) is not included in the model.

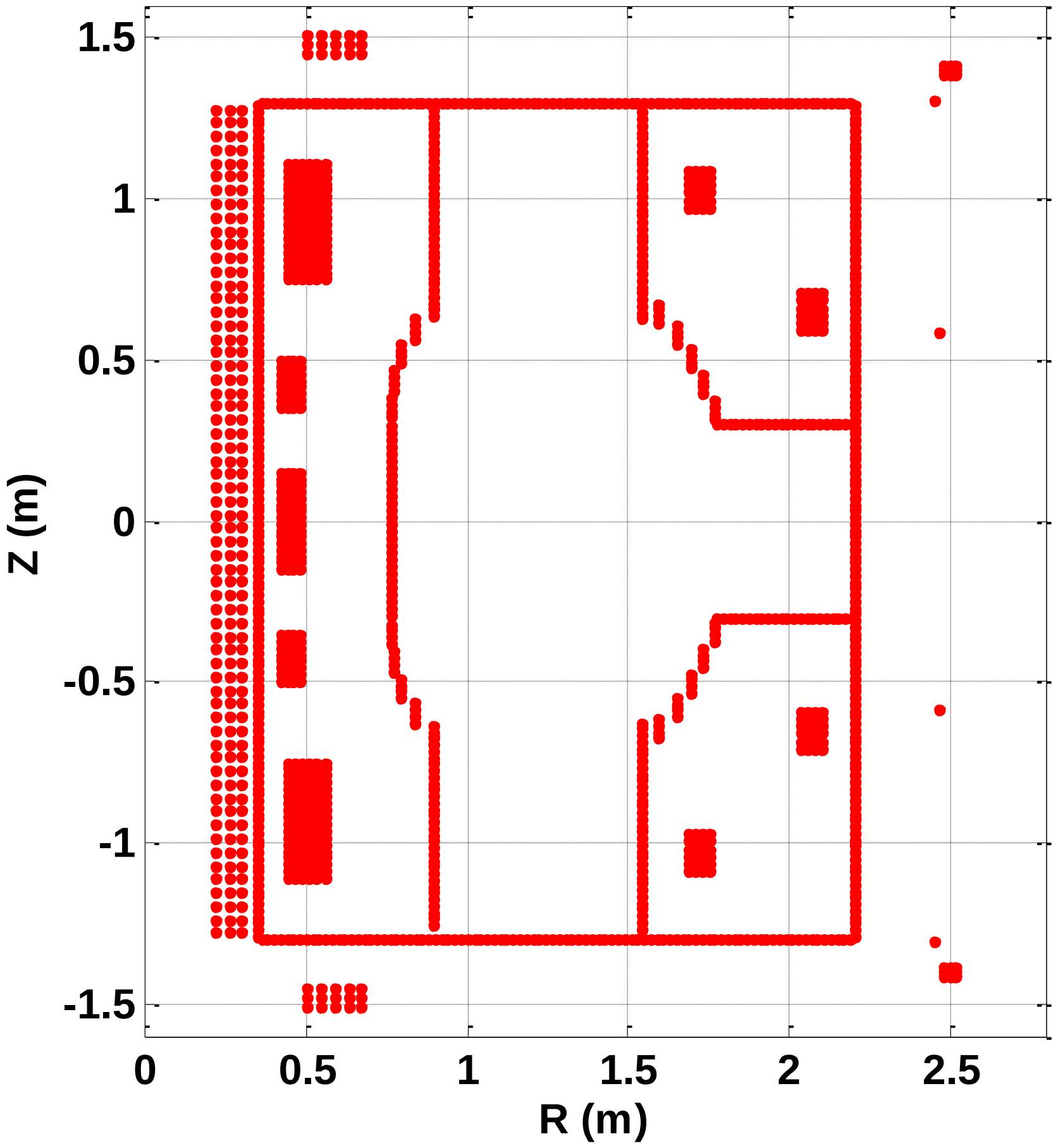

Fig. 1: Filament model of SST-1 prior to the installation of the plasma facing components. Each red dot represents the poloidal footprint of a toroidal filament.

The self-inductance for a circular loop conductor of radius *R* and cross-sectional area *A* with *N* turns is given by:

$$L = \mu_0 N^2 R \left[ \ln\left( \frac{8R}{\sqrt{A/\pi}} \right) - 1.75 \right] \text{ and} \tag{1}$$

The mutual inductance between two circular loops is given by:

$$M_{12} = N_1 N_2 \mu_0 \sqrt{ab} \left[ \left( \frac{2}{k} - k \right) K(k) - \frac{2}{k} E(k) \right] \tag{2}$$

Where, Where $N_1$ and $N_2$ are the number of turns and *a* and *b* are the radii of the first and second loop respectively. $K(k)$ and $E(k)$ are the complete elliptic integrals of the first and second kind respectively and *k* is elliptical modulus ($k^2 = 4ab/[(a+b)^2+(z_1-z_2)^2]$). Vertical positions (*Z* direction) of the first and second loop are given by $z_1$ and $z_2$ respectively.

Electromagnetic circuits are coupled via magnetic flux linkage and mutual inductance. The transformer action to induce the plasma current is provided primarily by a solenoidal Ohmic heating (OH), or induction, coil. A changing current in this solenoidal OH coil provides a changing magnetic flux through the surface bound by the plasma ring and hence the plasma current is built up. Similarly, induced current can be produced in the toroidal filaments. According to Faraday's law, a change in the magnetic flux through a surface bound by a circuit induces an electromotive force (voltage) proportional to $-d\Phi/dt$ is given by:

$$L_i \frac{dI_i}{dt} + R_i I_i = -\sum_{j \neq i}^{N} \frac{d\left(M_{ji} I_j\right)}{dt} \qquad i = 1, \ldots, N \tag{3}$$

The above ODEs govern the dynamics of $N$ coupled circuits. $N$ represents the number of toroidal filaments considered in the model. $I_i$, $L_i$, $R_i$ are the current, self-inductance and resistance of the $i^{th}$ passive filament. $M_{ji}$ is the mutual inductance between the $i^{th}$ passive filament and $j^{th}$ active filament. Note that the active filaments are for the active coils like OH, BV etc. $I_j$ is the current in the respective active filament. Eddy currents in all passive structures can be obtained by solving eqn. (3).

Table 1: Coil parameters in SST-1

| **No.** | ***R*** **(m)** | ***Z*** **(m)** | ***dR*** **(m)** | ***dZ*** **(m)** | $N_{turn}$ | $n_{fR}$ | $n_{fZ}$ | ***L*** **(H)** | ***Res*** **(Ohms)** |
|---|---|---|---|---|---|---|---|---|---|
| 1 | 0.26 | 0 | 0.12 | 2.6 | 660 | 3 | 62 | 0.035 | 0.0684 |
| 2 | 0.588 | 1.48 | 0.207 | 0.096 | 40 | 5 | 3 | 0.0027 | 0.0089 |
| 3 | 0.588 | -1.48 | 0.207 | 0.096 | 40 | 5 | 3 | 0.0027 | 0.0089 |
| 4 | 2.449 | 1.305 | 0.06 | 0.0238 | 3 | 1 | 1 | 0.0001 | 0.0029 |
| 5 | 2.449 | -1.305 | 0.06 | 0.0238 | 3 | 1 | 1 | 0.0001 | 0.0029 |
| 6 | 2.469 | 0.588 | 0.019 | 0.0227 | 1 | 1 | 1 | $1.76\times10^{-5}$ | 0.0011 |
| 7 | 2.469 | -0.588 | 0.019 | 0.0227 | 1 | 1 | 1 | $1.76\times10^{-5}$ | 0.0011 |
| 8 | 2.5 | 1.4 | 0.05 | 0.05 | 22 | 3 | 3 | 0.0073 | 0.0912 |
| 9 | 2.5 | -1.4 | 0.05 | 0.05 | 22 | 3 | 3 | 0.0073 | 0.0912 |
| 10 | 0.45 | 0 | 0.071 | 0.32 | 80 | 4 | 16 | 0.0064 | 1.5 |
| 11 | 0.45 | 0.425 | 0.071 | 0.163 | 40 | 4 | 8 | 0.002 | 1.5 |
| 12 | 0.45 | -0.425 | 0.071 | 0.163 | 40 | 4 | 8 | 0.002 | 1.5 |
| 13 | 0.5 | 0.93 | 0.136 | 0.383 | 192 | 6 | 18 | 0.0725 | 1.5 |
| 14 | 0.5 | -0.93 | 0.136 | 0.383 | 192 | 6 | 18 | 0.0725 | 1.5 |
| 15 | 1.72 | 1.03 | 0.085 | 0.136 | 40 | 4 | 6 | 0.0126 | 1.5 |
| 16 | 1.72 | -1.03 | 0.085 | 0.136 | 40 | 4 | 6 | 0.0126 | 1.5 |
| 17 | 2.07 | 0.65 | 0.085 | 0.136 | 40 | 4 | 6 | 0.0159 | 1.5 |
| 18 | 2.07 | -0.65 | 0.085 | 0.136 | 40 | 4 | 6 | 0.0159 | 1.5 |
| 19 | 1.766 | 0.344 | 0.01 | 0.08 | 1 | 1 | 4 | $1.00\times10^{-5}$ | 0.0499 |

The induced loop voltage ($V_i^L$) at the physical location of the $i^{th}$ flux loop ($Ri^F$, $Z_i^F$) is given by:

$$V_i^L = -\sum_{j\neq i}^{N} d\left(M_{ji} I_j\right)/dt - M_{iOH}\, dI_{OH}/dt - M_{iBV}\, dI_{BV}/dt \tag{4}$$

Where, $M_{ji}$ is the mutual inductance between $i^{th}$ flux loop and $j^{th}$ passive filament. $I_j$ and $R_j$ are the induced eddy current and resistance in $j^{th}$ passive filament respectively. $M_{iOH}$ and $M_{iBV}$ are the mutual inductances between $i^{th}$ flux loop and Ohmic coils and vertical field coils respectively. $I_{OH}$ and $I_{BV}$ are the conductor currents in Ohmic coils and vertical field coils respectively. Eqn. (4) is valid for vacuum shots in the absence of plasma current ($I_p$). In case of generated plasma current, another term ($-d(M_{pi}I_p)/dt$) will be added on the right hand side of Eqn. (4).

Tokamaks are equipped with small magnetic pick up coils installed on the vessel inner wall at tangential and normal angles to the vessel contour. These magnetic probes (MPs) can measure the fluctuations of the normal ($B_i^n$) or the tangential ($B_i^t$) component of a poloidal magnetic field. This measured magnetic field components can be compared with the output from the code as well. The geometrical center of the $i^{th}$ MP is denoted by ($R_i^B$,$Z_i^B$) ($1 \leq I \leq N_B$), where $N_B$ is the total number of MPs. We define $\alpha_i^B$ as the tangential angle of the $i^{th}$ MP with respect to the local vertical axis fixed at the geometric center of the MP, as shown in Fig. 9 (right panel). Then, $B_i^n$ and $B_i^t$ component of a poloidal magnetic field is easily computed from the simulation as,

$$B_i^t = B_i^R \cos\left(\alpha_i^B\right) + B_i^Z \sin\left(\alpha_i^B\right) \tag{5}$$

$$B_i^n = -B_i^R \sin\left(\alpha_i^B\right) + B_i^Z \cos\left(\alpha_i^B\right) \tag{6}$$

where, $B_i^R$ and $B_i^Z$ are the $R$ and $Z$ components of a poloidal magnetic field and given by:

$$B_i^R = -\frac{1}{R}\frac{\partial\psi}{\partial Z},\ B_i^Z = \frac{1}{R}\frac{\partial\psi}{\partial R} \tag{7}$$

A toroidal wire loop at ($R_s$,$Z_s$) with current $I$ produces the flux $\psi(R,Z)$ in ($R$,$Z$):

$$\psi\left(R,Z\right) = \frac{\mu_0}{\pi} IG \tag{8}$$

where, $G(R,Z,R_s,Z_s)$ is the Green's function [10], defined as $G=M_{12}/2\mu_0 N_1 N_2$.

It can be noted that the active coils, like TR and BV, are also broken into a set of filaments distributed according to the radial (*R*) and vertical (*Z*) spans of the respective coils in this code. This makes the mutual inductance matrix calculation even more accurate due to the physical dimensions of the respective coils. The number of filaments along *R* and *Z* directions are given as $n_{fR}$ and $n_{fZ}$ in table 1.

## IV. Experimental verification

### A. *With toroidal flux loops*

SST-1 is equipped with 23 flux loops, 11 of them are installed inside the vacuum vessel, 2 are installed inside the cryostat and the rest 10 are installed outside the cryostat. A cross-sectional view of SST-1 with the poloidal field (PF) coils and the installed flux loops are shown in Fig. 2.

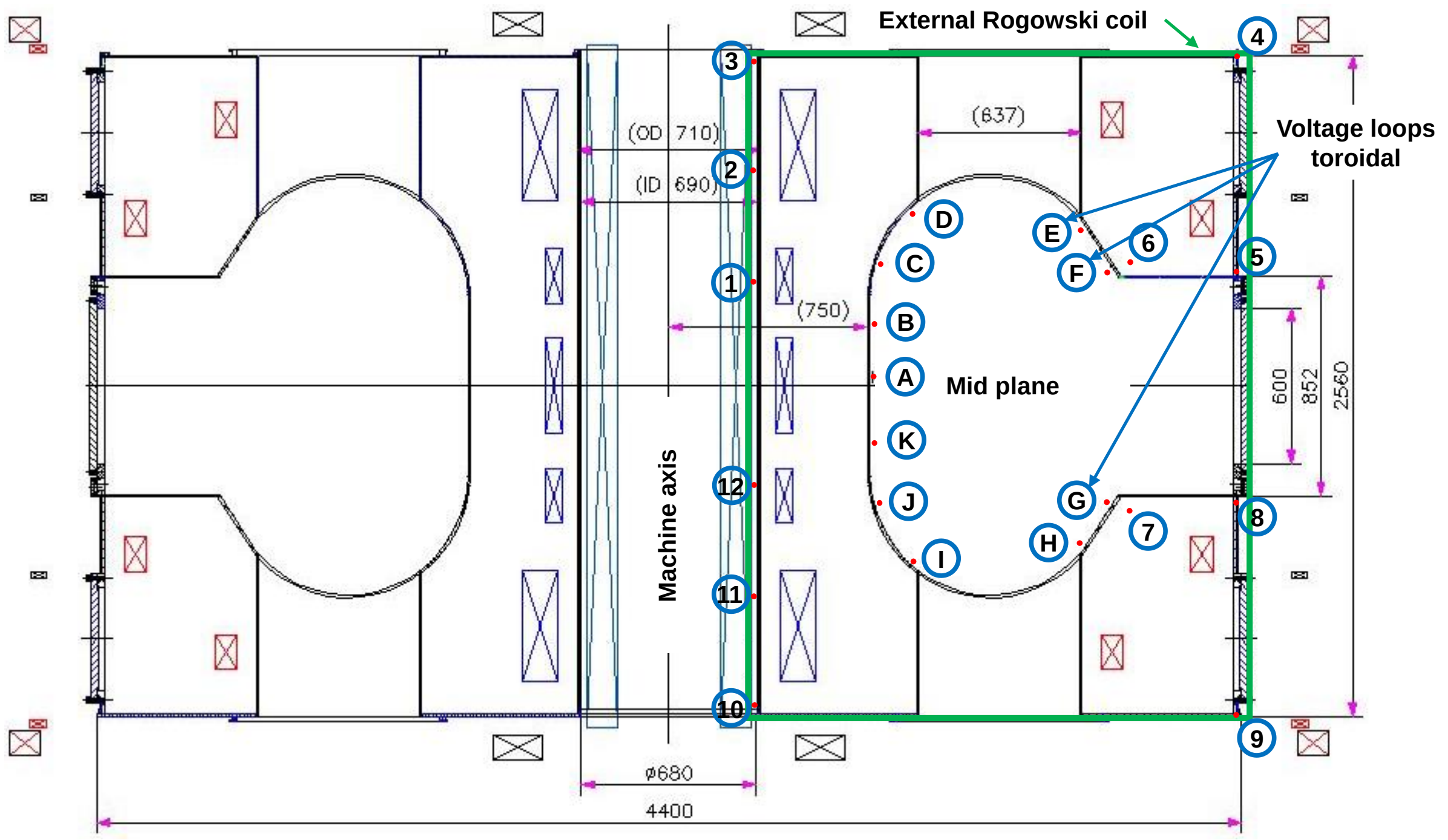

Fig. 2: Cross-sectional view of SST-1 with the PF coils and Ohmic transformer sections shown. Also 11 in-vessel (A-K) and 12 (1-12) out-vessel flux loops can be seen. The solid green line represents the proposed external Rogowski coil, shown by the green arrow. Dimensions are in mm.

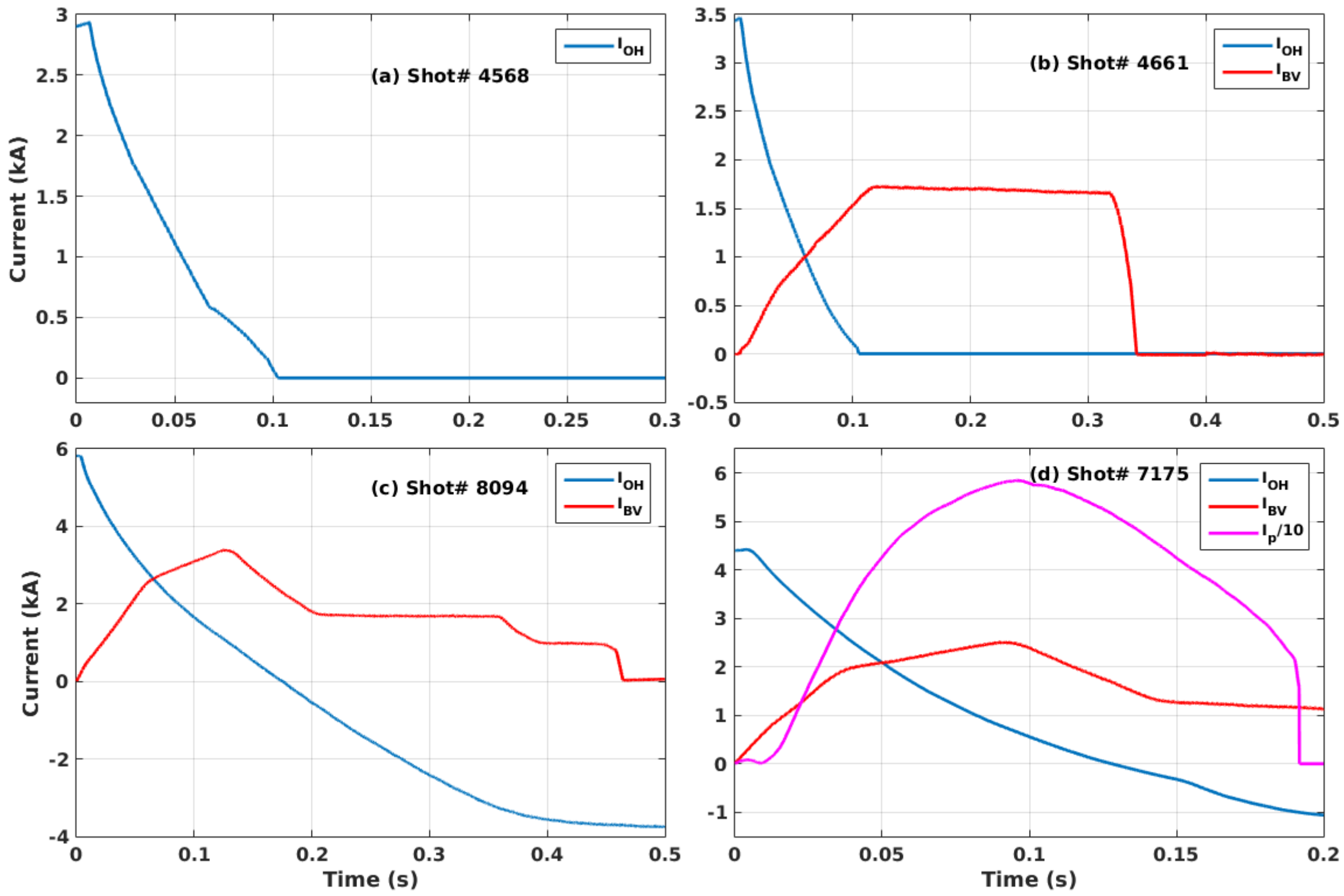

Fig 3: Discharge waveforms for four shots.

Once the filaments currents are calculated for a given OH coil current ($I_{OH}$), loop voltage due to this filaments at the physical locations of the installed flux loops can be estimated. This can be done with and without a vertical magnetic field (BV) used for sustaining the plasma. Fig. 3(a) shows the waveform of OH coil for the representative shot# 4568. The estimated and measured loop voltages at all the 23 flux loop locations are compared. Fig. 4 and 5 show the comparison of measured and estimated loop voltages for the in-vessel and out-vessel flux loops for this shot, taken with the OH coil contribution only. The same exercise is carried out for another shot with both the OH coil and BV contributions together. Fig. 6 shows the $I_{OH}$ and $I_{BV}$ waveforms. Measured and simulated loop voltages are shown in Fig. 7 and 8. Excellent agreement between the measured and simulated loop voltages has been achieved for most of the flux loops except the loops designated as B and 10. These two loops are having some hardware related problems and will be sorted out in the future vacuum vessel opening schedules. Fig. 7 and 8 also show the individual contribution to loop voltages due to $I_{OH}$ (magenta) and $I_{BV}$ (green) separately and together (black), in absence of all the passive elements and their corresponding filaments. It can be seen that the passive elements are introducing a considerable delay as recorded in the loop voltages in all the in-vessel and low field side out-vessel flux loops. The effect of passive elements (filaments) are minimal on the loops mounted on the center stack. Magnitude of the introduced delay increases as one moves away from the center stack.

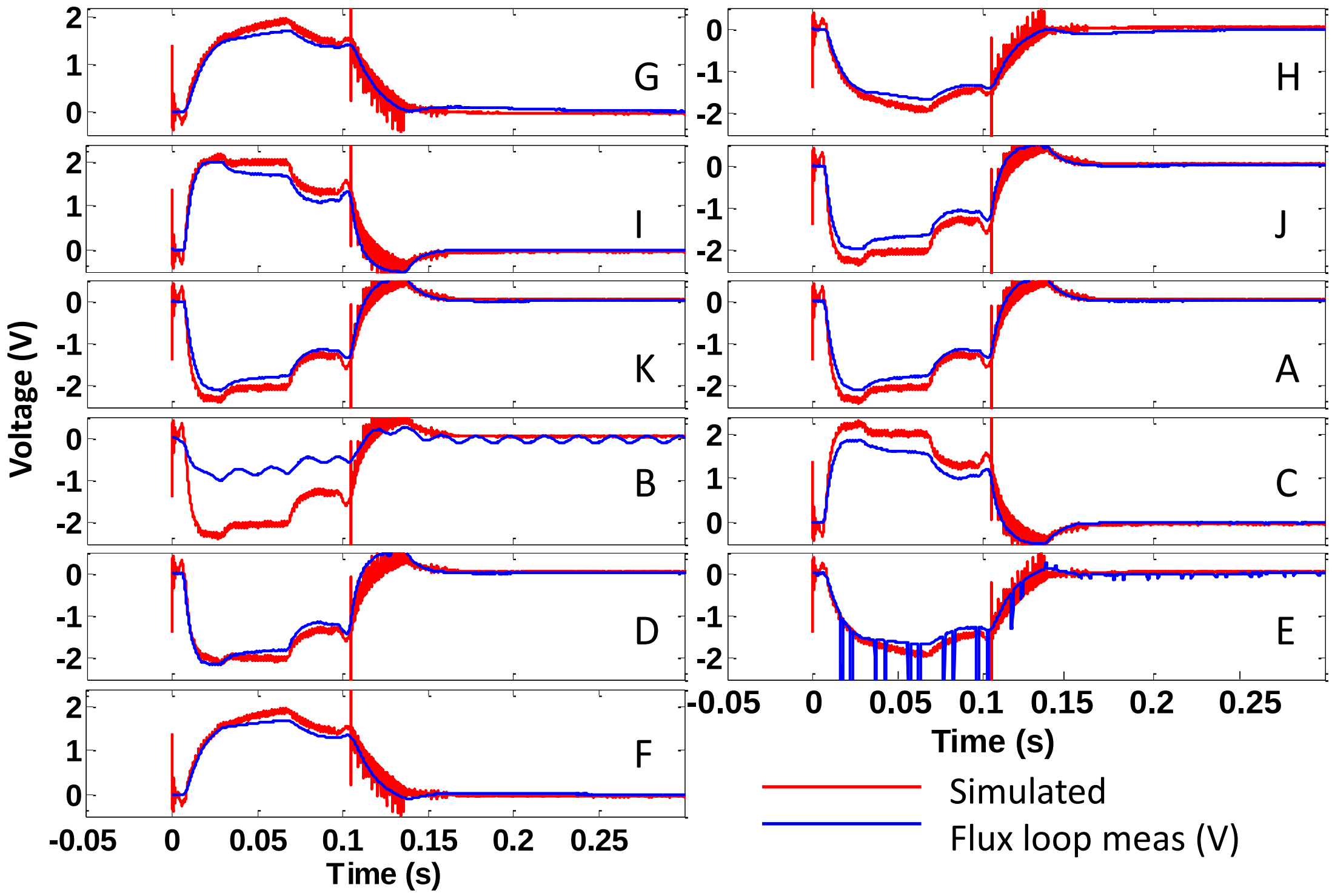

Fig. 4: Vacuum shot #4568 – Pure OH; In-vessel flux loops

It can be seen from Fig. 6 that after ~165 ms, slope in $I_{OH}$ becomes negligible. Thus loop voltage signals also attain values ~0 for the shot #4364 (refer Fig. 7 and Fig. 8). However, after ~200 ms there is a zero cross-over in loop voltages. This part is purely dominated by the negative slope in the $I_{BV}$, as evident from Fig. 6. It can be noted that $I_{OH}$ and $I_{BV}$ are of opposite polarity for a typical tokamak discharge. The zero cross-over part is shown in Fig. 7 and 8 to illustrate the sensitivity of the flux loops and the accuracy of our simulation. A small dip in $I_{BV}$ is also captured nicely in the green plots (with $I_{BV}$ only without the passive elements' contribution) at ~110 ms. For actual plasma shots in SST-1, sufficiently long pulse of $I_{BV}$ will be given to sustain the plasma for the desired duration. This procedure helps as an essential benchmarking for the eddy current modeling and demonstrates the sensitivity of the approach.

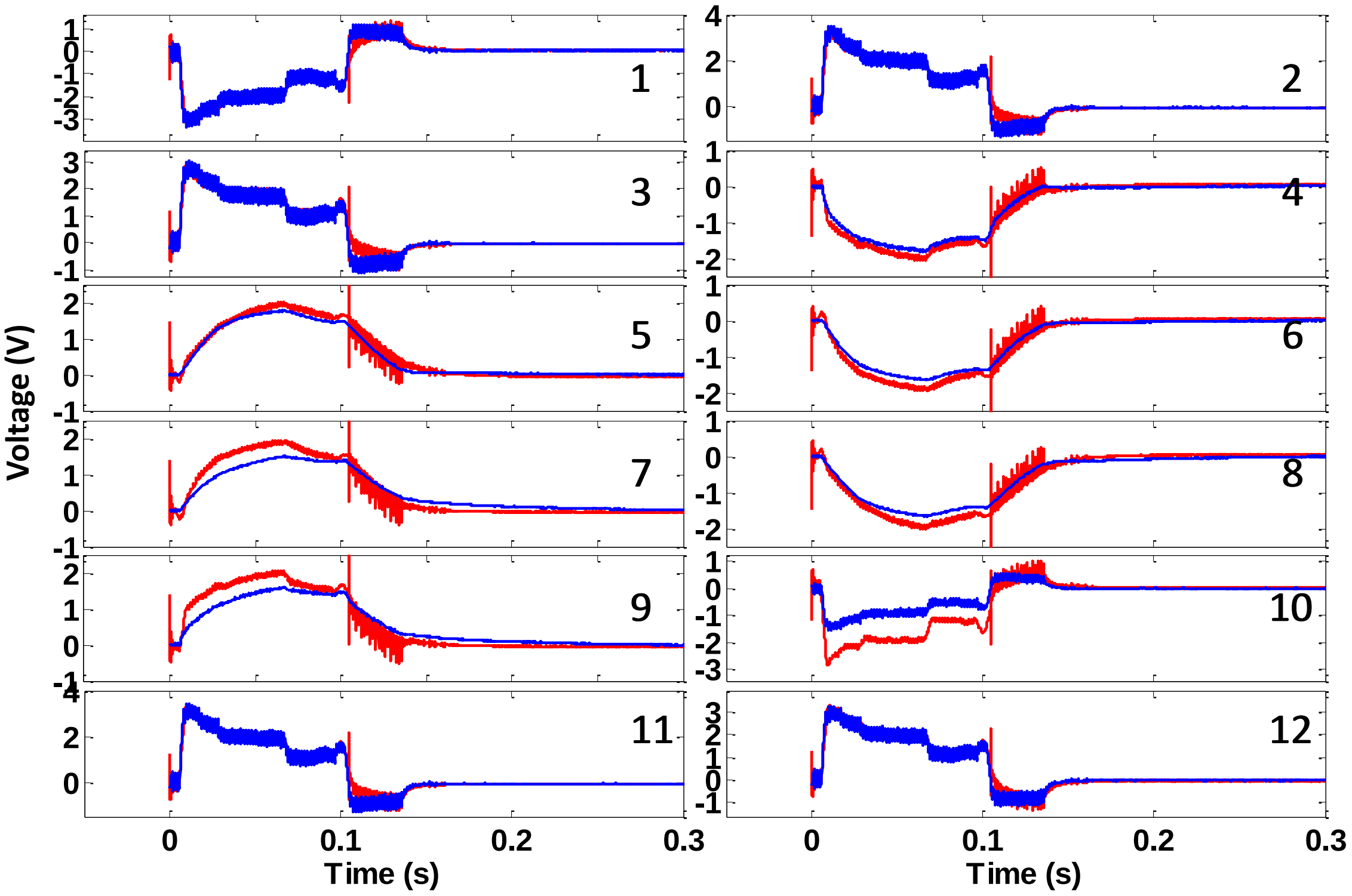

Fig. 5: Vacuum shot #4568 – Pure OH; Out-vessel flux loops

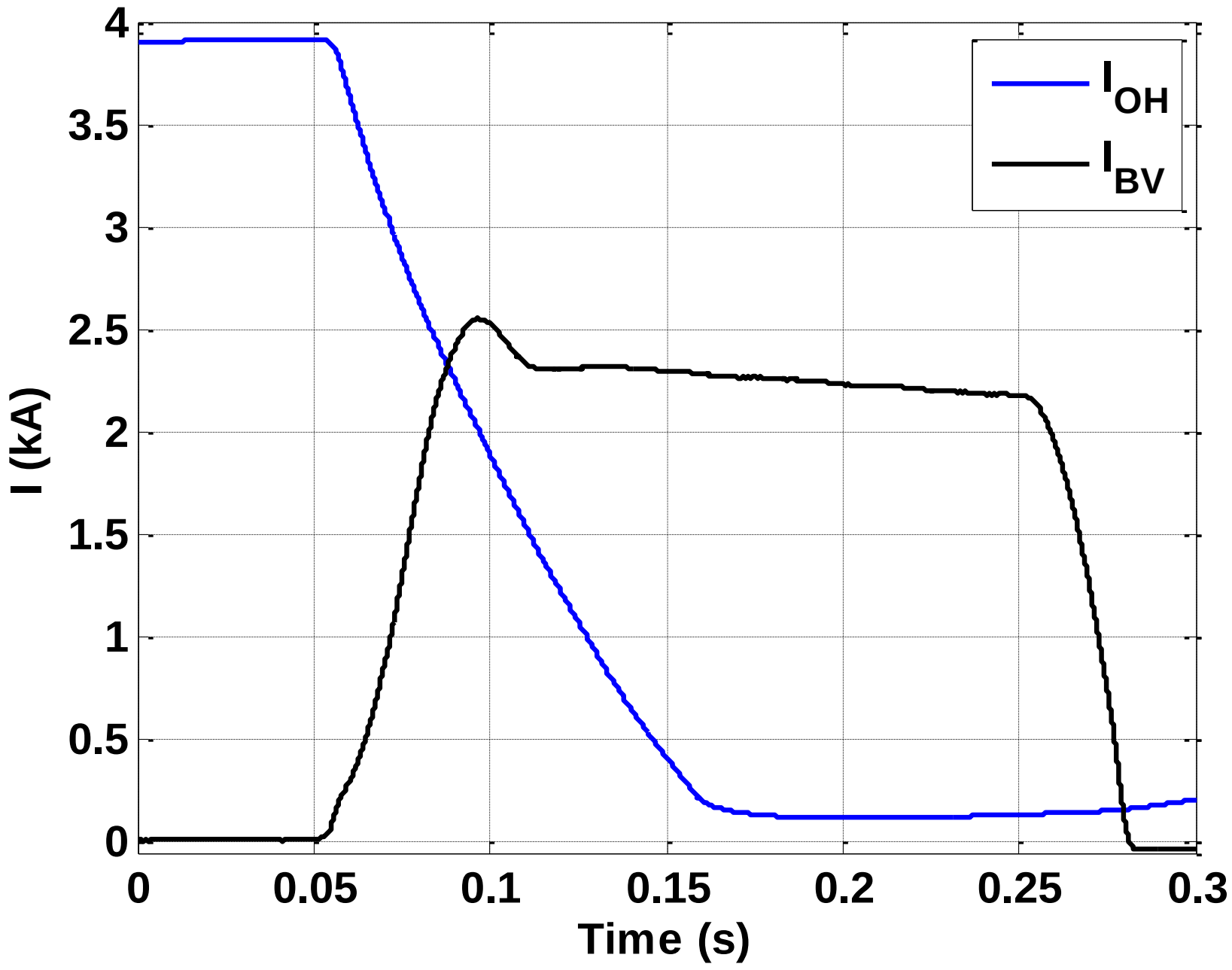

Fig. 6: Vacuum shot #4364 – OH+BV; Wave forms of $I_{OH}$ and $I_{BV}$ are shown respectively.

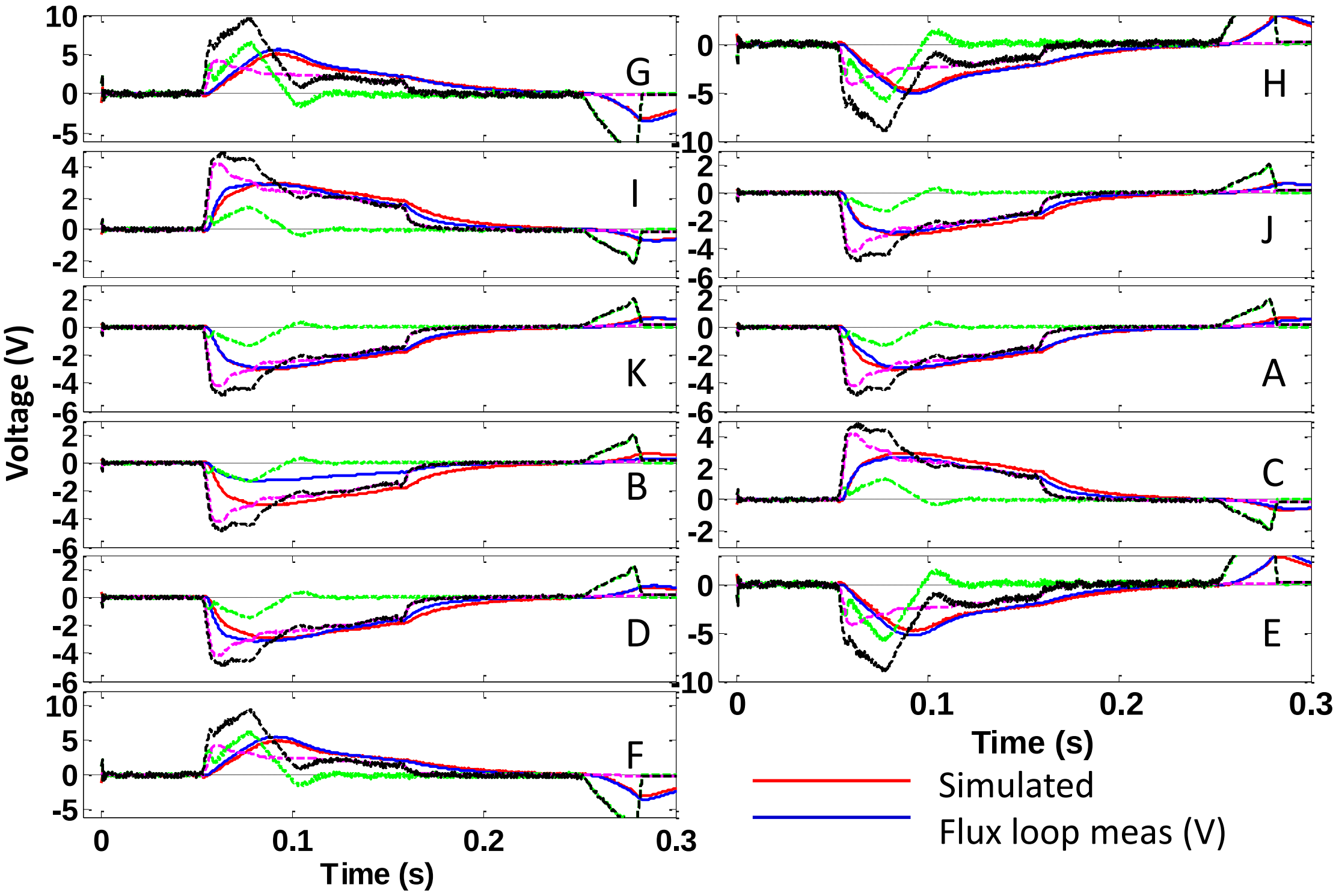

Fig. 7: Vacuum shot #4364 –OH+BV; In-vessel flux loops

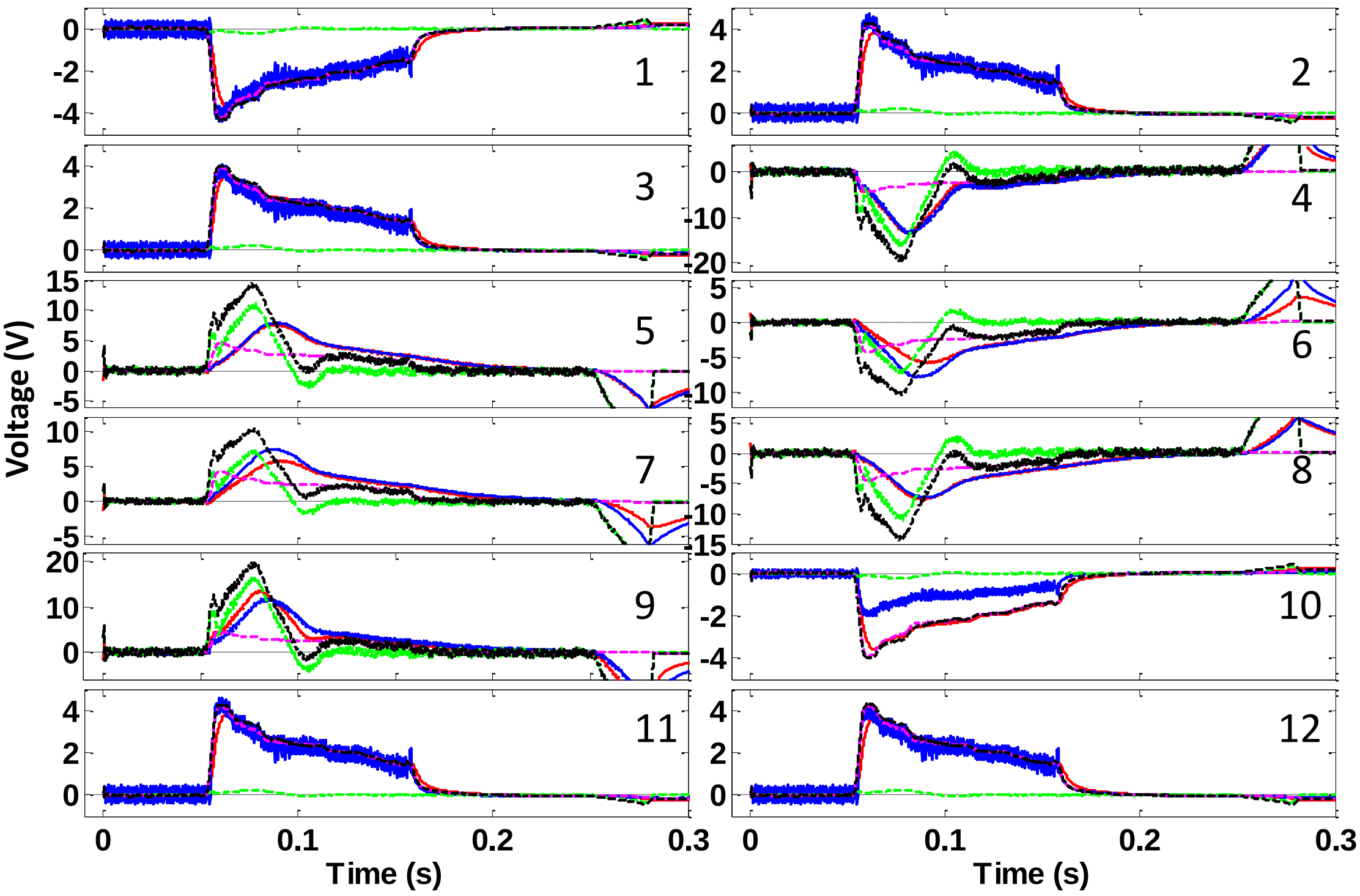

Fig. 8: Vacuum shot #4364 –OH+BV; Out-vessel flux loops

## B. *With position probes*

Apart from the flux loops, SST-1 is also equipped with magnetic pick-up coils (position probes). The tangential set of coils (‘T’ coils) are comprised of 12 coils arranged on the vacuum vessel wall such that they follow the poloidal contour of the vessel. These coils are tangential to the poloidal cross-section of the vessel. The amplifier section of the probes are equipped with a 1 kHz filter to block high frequency part as these signals are mainly intended for generating the plasma position information. Data is acquired at 100 kHz. The design value of the effective area $A_{eff}$ for the probes is 0.044 turn-m$^2$. The next step is to compare the poloidal magnetic field output of the filament model with the $B$ signals derived from the measured $\dot{B}$ signals. This adds both redundancy and rigor to the estimated filament currents. The coils positions and their legends are shown in Fig. 9 (left panel), while their angles with the local vertical axes are shown the right panel. Experimentally acquired $\dot{B}$ signals are first normalized with the applied gain in the acquisition electronics and then offsets (if any) have been subtracted. This pre-processed signal is then software integrated to generate the measured poloidal magnetic fields. Raw $\dot{B}$ signals acquired during a vacuum shot #4661 are shown in Fig. 10. Discharge waveforms for OH and BV coils for this shot are shown in Fig. 3(b). The probes which are at 90 (Pr3) and 270 (Pr10) degrees respectively with the mid-plane are not expected to pick up much flux from the OH and $B_v$ coils. However, these probes still show signal in vacuum shots due to some physical inaccuracies in the installation. Pr12 was having some hardware problem in this campaign and the acquired signal is a bit noisy. Comparison of the experimental and simulated profiles of the $B$ signals for all the probes are shown in Fig. 11. Here the experimental signals are compared with the simulated signal by multiplying a constant factor all across the waveform. This factor, in turn-m$^2$, is unique to the concerned position probe and depends on the effective area of the probe (a combination of the sensitivity achieved and the effective area realized for that probe after installation). It is statistically robust on a large number of shots with similar operating conditions. These factors are 0.03±0.01 for all the probes. Apart from Pr3 and Pr10, reasonably good agreement has been achieved with the experimental signals. As expected, waveforms from experiment and simulation for Pr3 and Pr10 do not match. It can be seen in both Fig. 10 and Fig. 11 that there are some pick-ups from the OH coil in the probes and that is significant in the initial phase. Variations between the experimental and simulated profiles in some of the probes can be attributed to the physical inaccuracies in the installation like orientation etc.

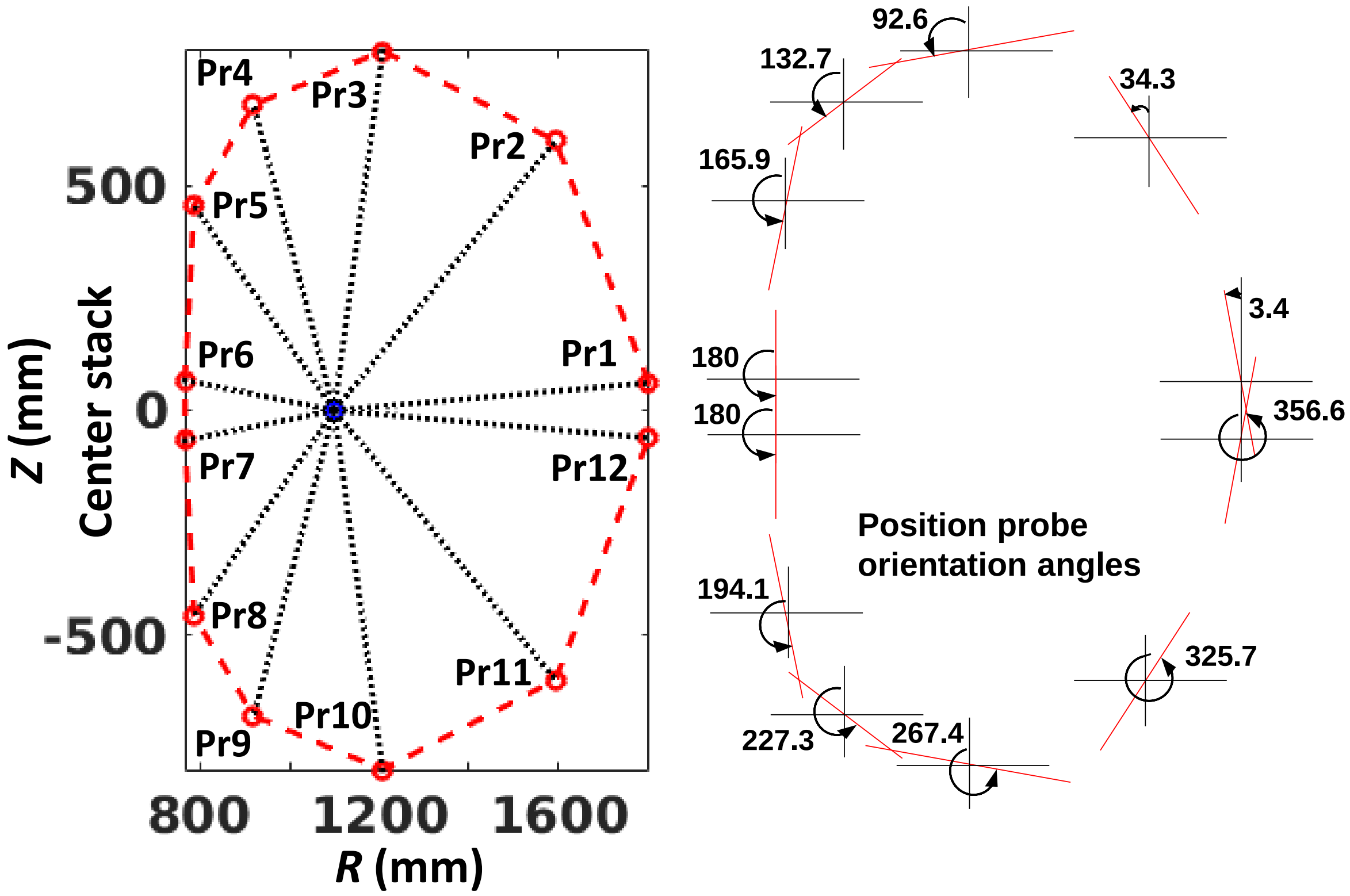

Fig. 9: Magnetic pick-up coils nomenclature and their distribution are shown as red circles on the SST-1 vacuum vessel cross-section. The orientation angles $\alpha_i^B$ for all the probe are shown on the right.

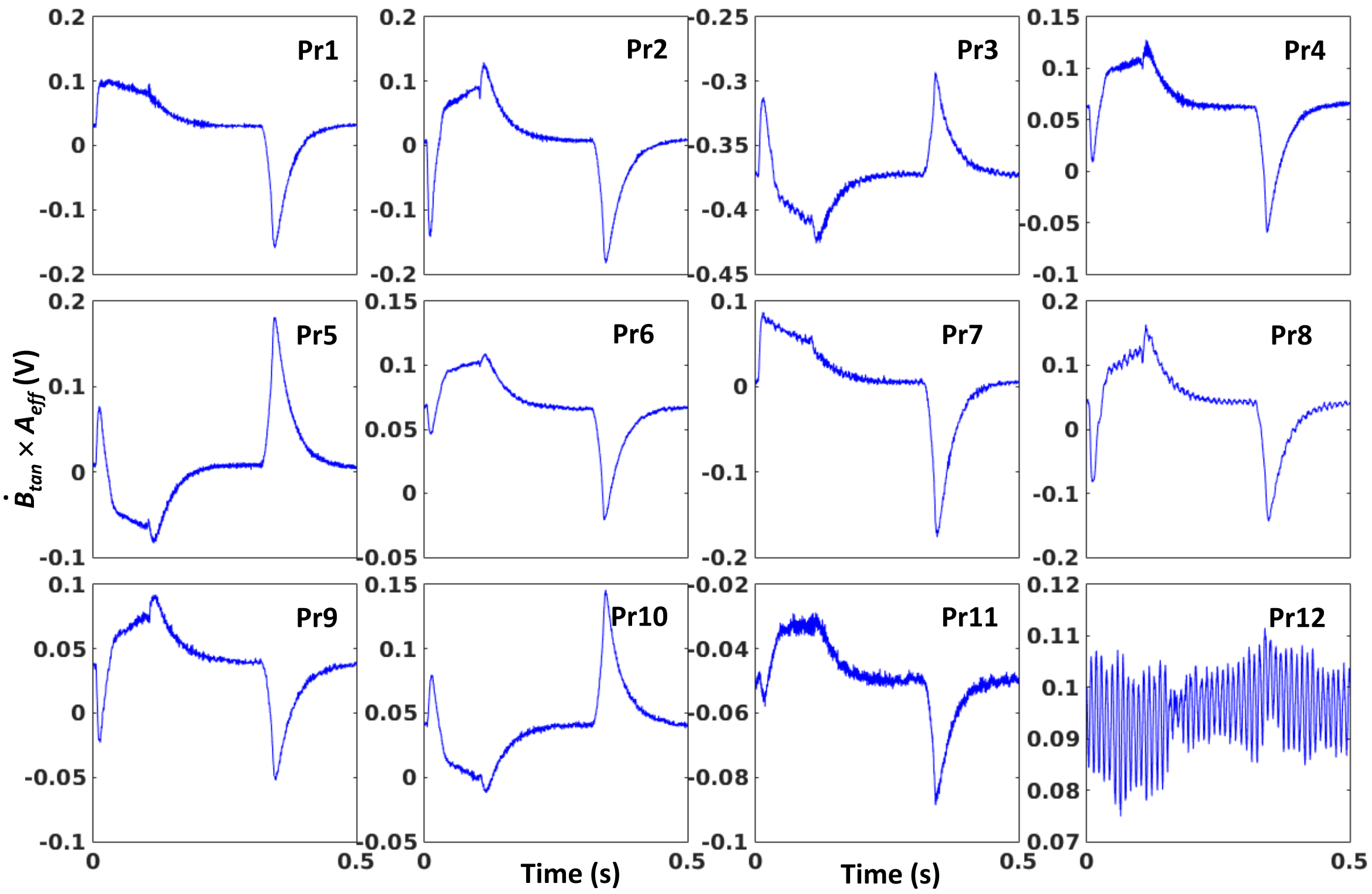

Fig. 10: Shot #4661: Raw signal of 12 tangential position probes that were giving good signals for these shots.

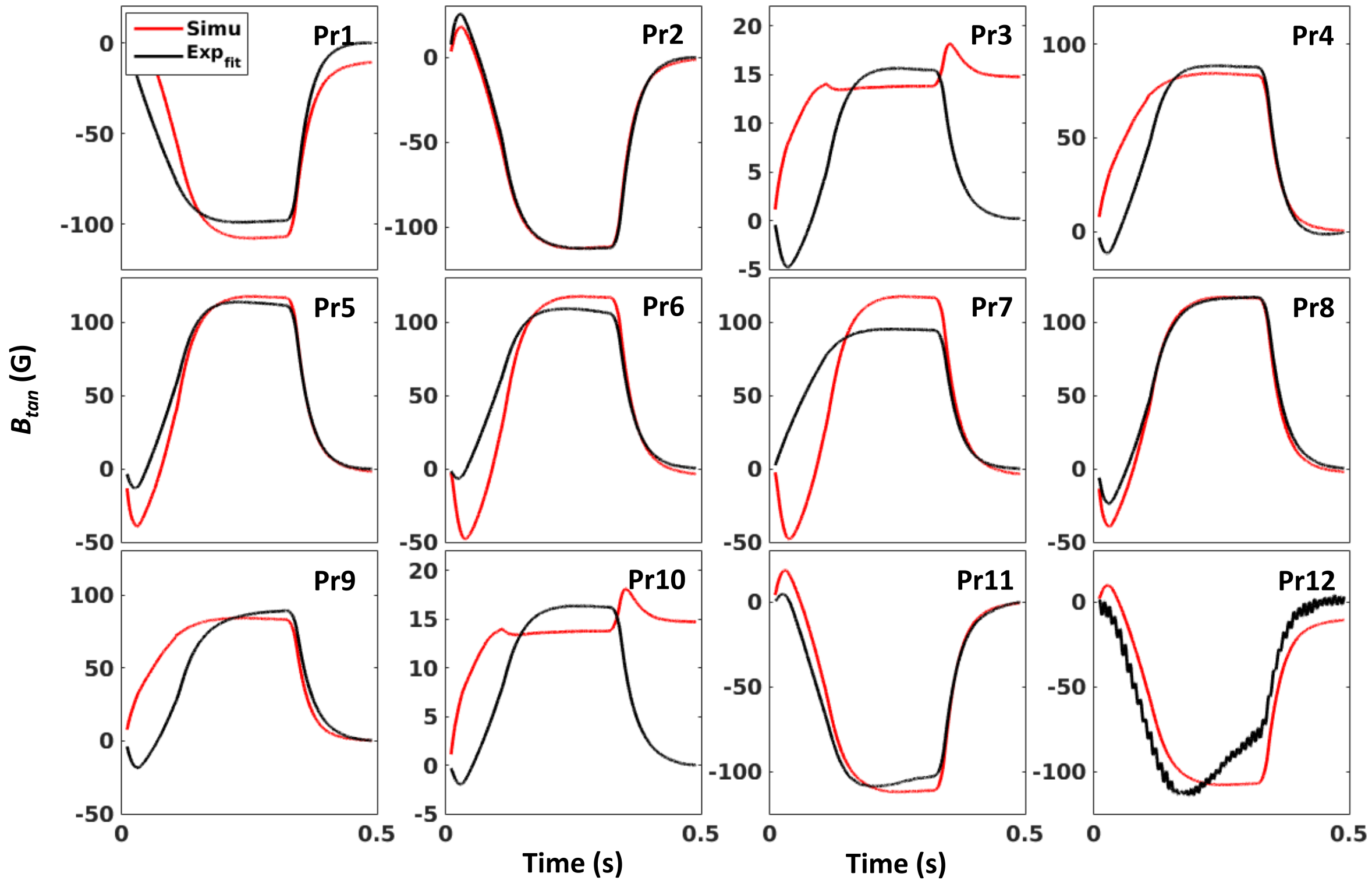

Fig. 11: Vacuum shot #4661; Simulated and experimental signals are compared for the 12 probes.

### C. *Filament model upgraded with the inclusion of PFCs*

Plasma facing components (PFC) are installed in SST-1 for phase II of operations [11]. These passive structures may introduce small corrections to the magnetic geometry, albeit they are in saddle-configuration on the SST-1 machine. Hence, the PFCs need to be included in the filament model to get a comprehensive understanding of the induced EMFs. Fig. 12 shows the filaments added to the existing model to include the PFCs. 100 new filaments are added to represent the PFCs. Out of these 100 filaments 21 pairs are in saddle configuration representing the outer passive (15 pairs) and inner passive (6 pairs) stabilizers respectively. These filaments are shown in blue in Fig. 12. Rest of the PFC filaments are toroidally continuous similar to the vessel/cryostat filaments and represent the divertor plates. Resistance of a saddle pair is simply determined by adding the individual resistances of the participating loops (top and bottom) in the pair and resistances for the legs, as shown in Fig. 13. The self-inductance for a saddle pair filament is realized as:

$$L_s = 2(L - M_{12}) \tag{9}$$

Where $L$ is the self-inductance of each loop of the saddle pair and $M_{12}$ is the mutual inductance between the top and bottom loops in a pair. The mutual inductance between two saddle pairs is realized as:

$$M_{12}^{s} = 2\left(M_{11'} - M_{12'}\right) \tag{10}$$

Where $M_{11'}$ is the mutual inductance between the top loops of first and second saddle pairs and $M_{12'}$ is the mutual inductance between the top loop of first and bottom loop of the second saddle pair. Prime stands for the second loop. Finally, mutual inductance between a saddle pair and toroidally continuous simple filament is realized as:

$$M^{st} = M_{11'} - M_{21'} \tag{11}$$

Where $M_{11'}$ is the mutual inductance between the top loop of the second saddle pair and toroidal filament and $M_{21'}$ is the mutual inductance between the bottom loop of the second saddle pairs and the toroidal filament. Self and mutual inductances are calculated as given earlier in equations (1) and (2) respectively.

Fig. 14 shows the comparison of experimental and simulated $B$ signals for all the 12 position probes for a vacuum shot (Shot #8094) after PFC installation. Discharge waveforms for OH and BV coils for this shot are shown in Fig. 3(c). As stated earlier Pr3 and Pr10 do not match well for this vacuum shot. Rest of the probes can be matched reasonably well considering the physical installation inaccuracies.

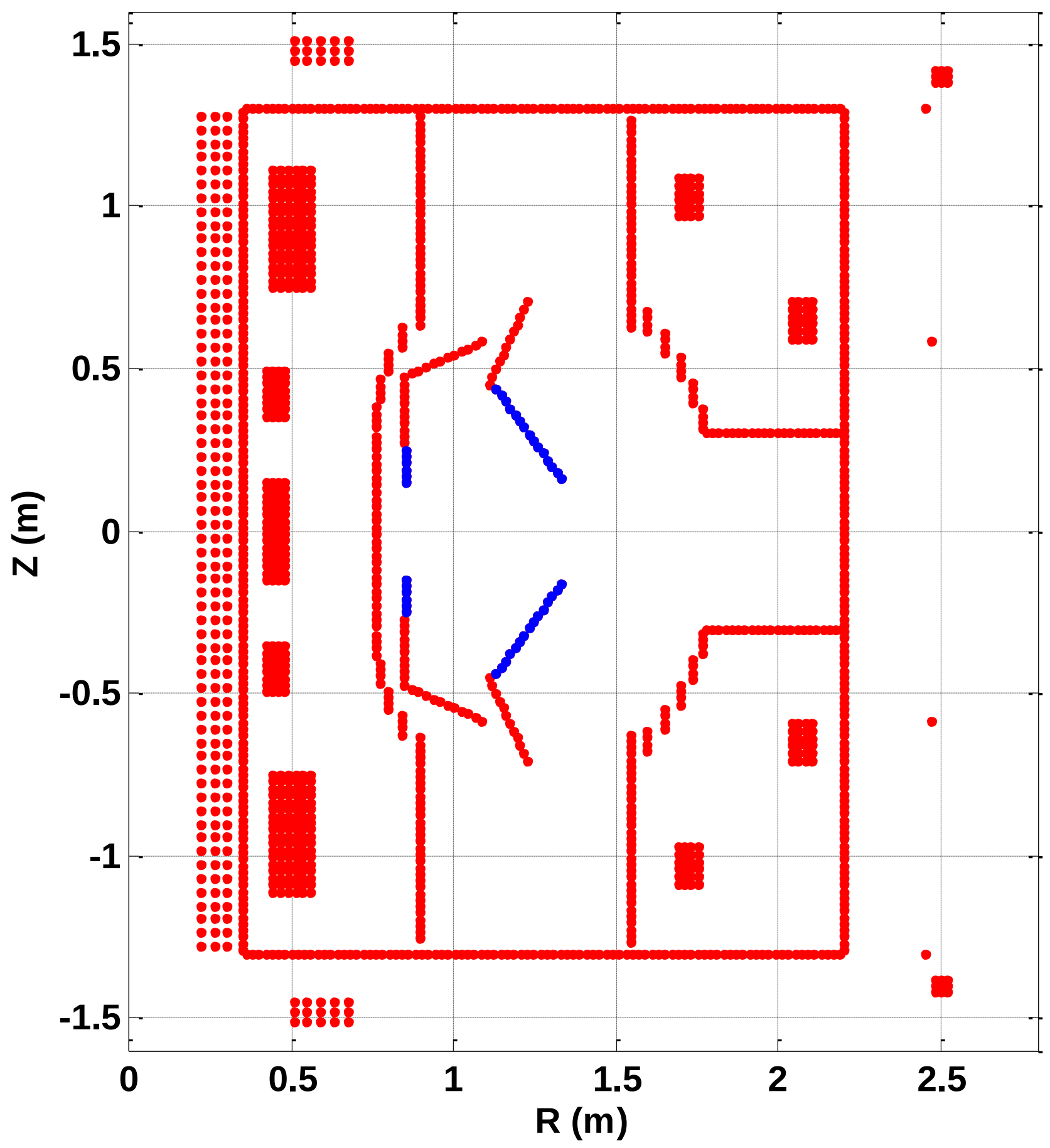

Fig. 12: Filament model of SST-1 including the plasma facing components. Each red dot represents the poloidal footprint of a toroidal filament as in Fig. 1. The blue dots are the filaments in saddle pair representing the inner and outer passive stabilizers respectively.

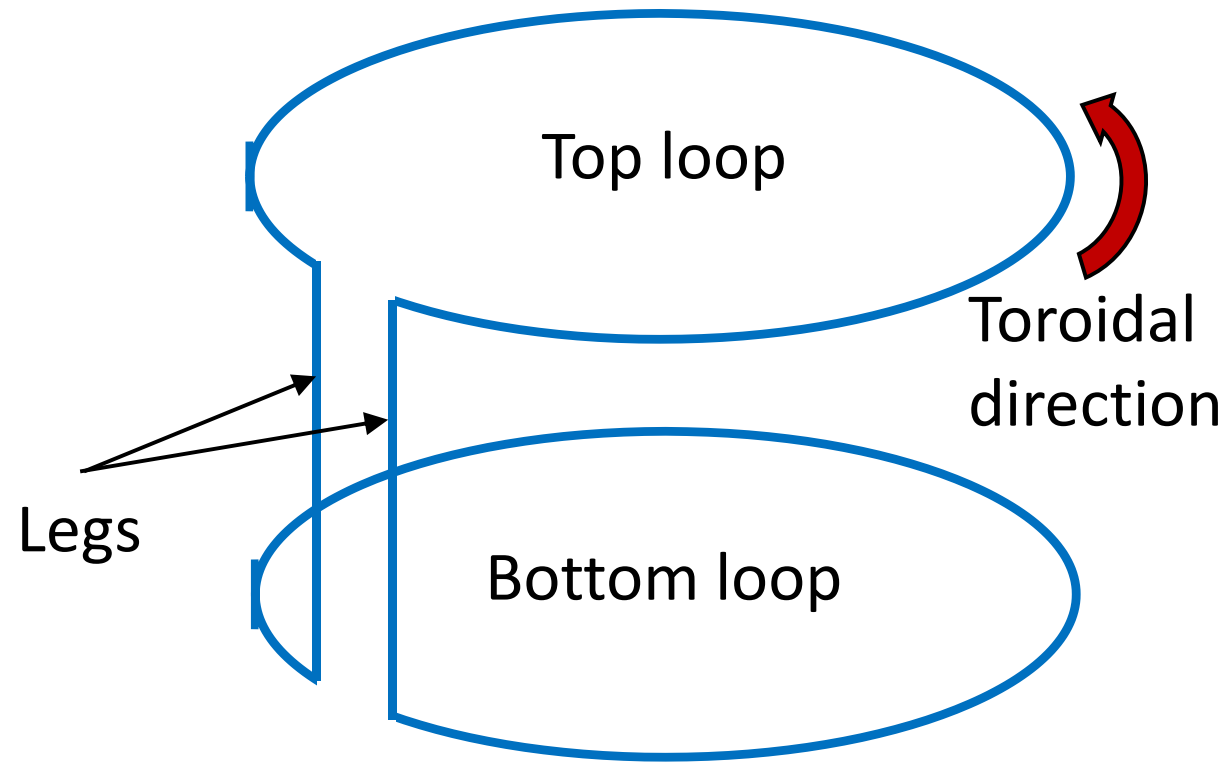

Fig. 13: Schematic of a saddle filament pair showing the top and bottom loops connected with the legs.

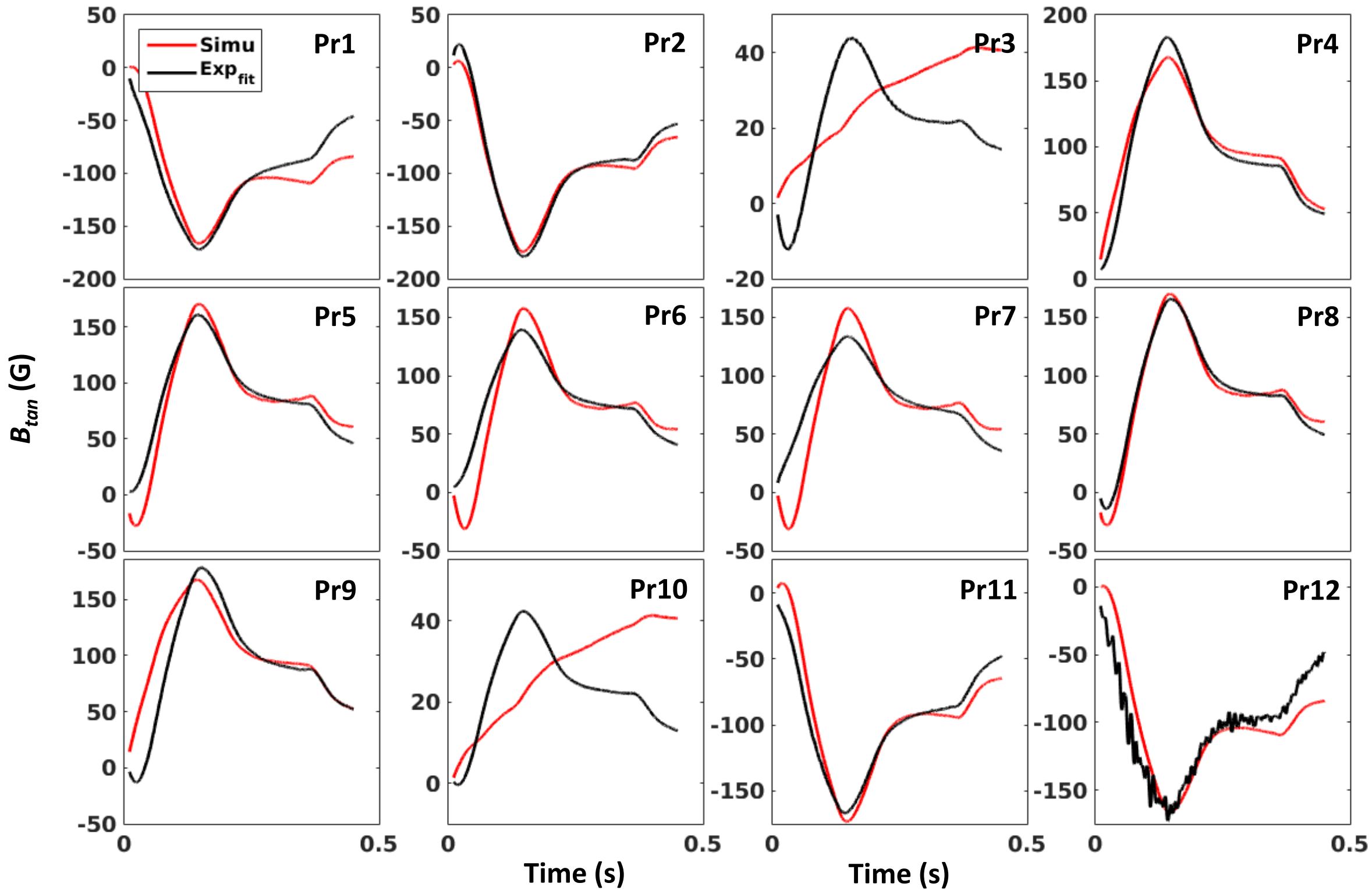

Fig. 14: Vacuum shot #8094; Simulated and experimental signals are compared for 12 probes.

The calculated eddy current in all the passive elements $I_{ext}$ are shown in Fig. 15. It can be seen that the currents flowing in the passive structures are quite significant and will play a major role in the plasma control and equilibrium. This total current will be measured by the recently installed external Rogowski coil as shown in Fig. 2 in the future campaigns of SST-1. It is noteworthy here that $I_{ext}$ has a steep slope in the first ~10 ms. However, the position probe signals do not show that feature. This is due to the fact that the conducting vacuum vessel and cryostat shields the high frequency (~0.1 kHz) part from penetrating to the position probe locations inside the vessel and also introduces a delay of ~13 ms in the waveforms. These effects were also seen in the plasma current ramp-up rate and field penetration in case of actual plasma shots and may pose serious challenges in plasma control.

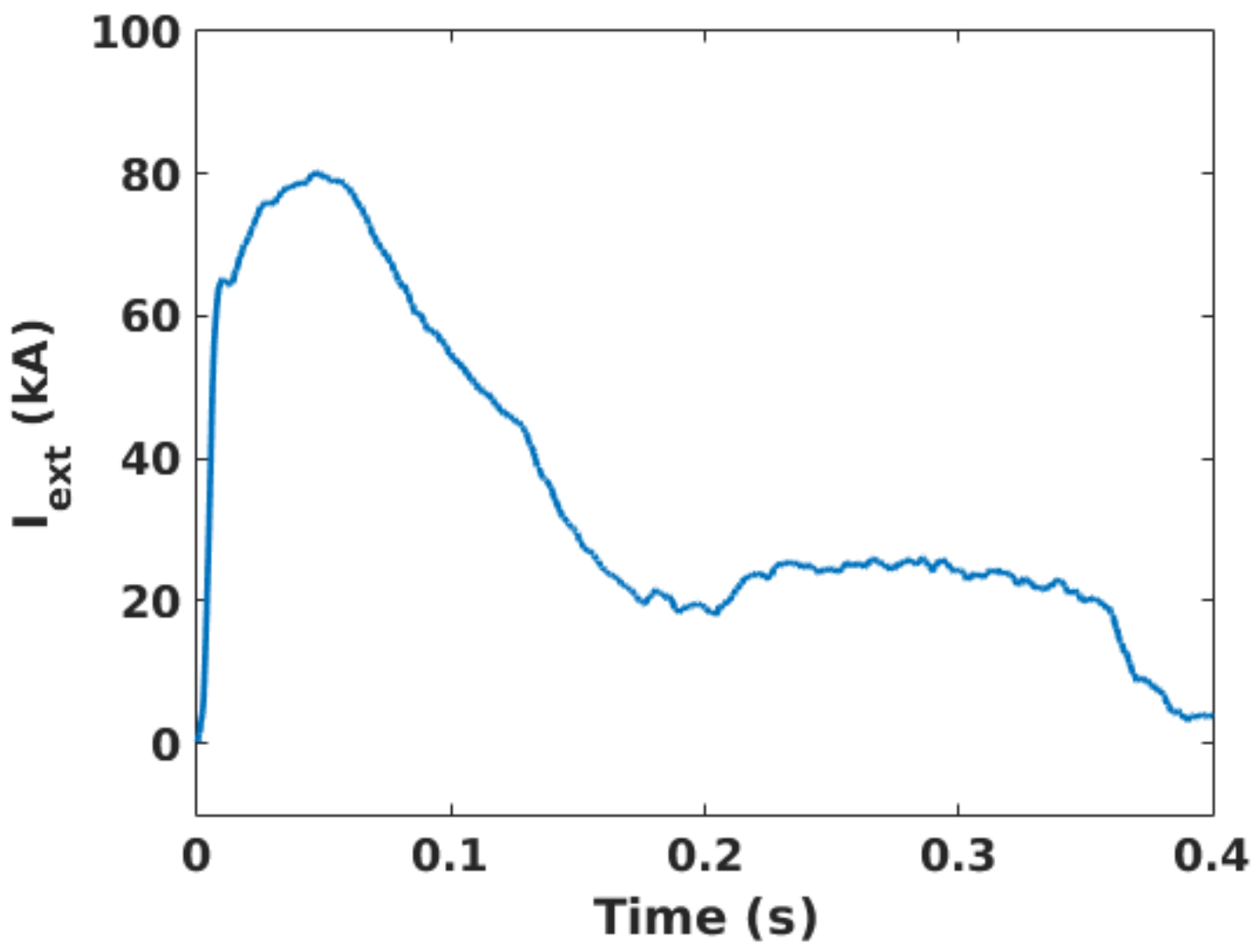

Fig. 15: Vacuum shot #8094; Total eddy current flowing in the passive structures

After analyzing the simulated and measured loop voltages and position probe signals in vacuum shots, we tried to match the experimental signals in the presence of plasma current. The plasma current ($I_p$) measured by the in-vessel Rogowski coil is replicated by a toroidal, single turn active current carrying conductor of square cross section (10 cm × 10 cm), placed at the plasma major radius. This $I_p$ channel replica is maintained at the $R$ = 1.1 m and $Z$ = 0 m all through the actual $I_p$ duration and is divided into a 5 × 5 toroidal filament matrix. Radial profile of the plasma current and radial shift of the plasma column during the course of evolution of the shot are not taken into account in this simplified representation at the moment and that will be finalized after the equilibrium reconstruction is carried out in future. Discharge waveforms for OH and BV coils and measured $I_p$ for this shot are shown in Fig. 3(d). Fig. 16 shows the comparison of the magnetic pick-up coils for a plasma shot. Reasonably good agreement of the experimental and simulated signals is obtained for all the probes except Pr12. There were some hardware related problems in the probe Pr12 in this set of shots. There are also some variations, like in probes Pr2, Pr3, Pr8, Pr9 and Pr10. This exercise is a test case example and at present only a simplified representation of $I_p$, as a central current carrying conductor, is demonstrated. It can be noted here that the $I_p$ channel is replicated as an active current carrying conductor here and hence not generated by the applied loop voltage. Also, Pr3 and Pr10 are now showing proper waveforms as the replicated $I_p$ channel is generating true poloidal field unlike the vacuum shots. The fitting will improve when a proper distribution of the plasma current will be considered. This exercise shows that the eddy current distribution can be modelled with reasonable precision even in the presence of $I_p$.

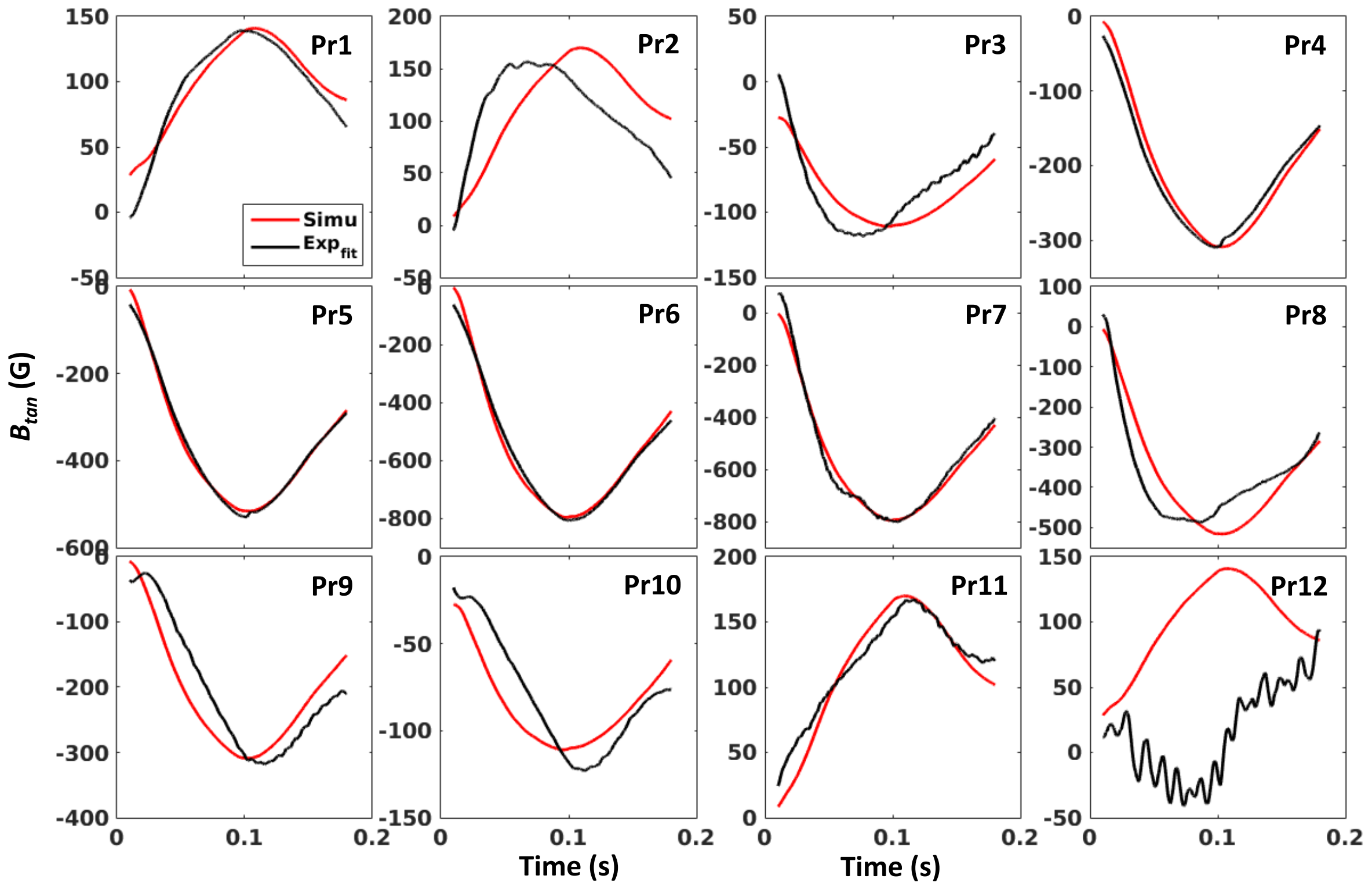

Fig. 16: Plasma shot #7175; Simulated and experimental signals are compared for all the 12 probes.

## V. Discussions

Induced eddy current distribution in the toroidal direction is modelled in SST-1 and validated against a number of experimental signals of varied types. These include both in-vessel and out-vessel toroidal flux loops and magnetic pick-up coils. Excellent agreement has been achieved for most of the measured signals. Hence, the resultant distribution should provide a realistic input to the equilibrium code. It can further be noted that the experimental signals from the probes with least errors will also be used in the minimizer for the equilibrium reconstruction code later on. The minimizer helps in reconstructing the equilibrium while minimizing the root mean squared error among the experimental and reconstructed signals in each iteration of the reconstruction algorithm.

The view ports and other complex mechanical structures are simplified in the model with toroidally symmetric filaments. This makes the model calculations fast and quite easy to handle. Nevertheless, the measured and simulated magnetic signals vary by <2% for most of the flux loops, as evident from Figs. 4, 5, 7, 8 and <5% for a large number of magnetic probes, as shown in Fig. 14. Hence, the axisymmetric model can replicate the actual scenario to an appreciable extent and should suffice for the further studies of plasma startup, break down, control, equilibrium etc. In future, this model will be further validated against 3D calculations using the commercial FEM software like COMSOL® Muliphysics [5].

## Acknowledgements

Contributions from the support staff of SST-1 tokamak are gratefully acknowledged. The authors would like to acknowledge K. R. Vasava for providing the CAD drawings. One of the authors (SB) would also like to acknowledge J. Ghosh, S. Pradhan, P.K. Kaw, A. Das and K. Mishra for many fruitful discussions during the course of this work.